\documentclass[pra,showpacs,graphics,twocolumn,floatfix,mathbbm,a4paper]{revtex4-1}
\usepackage{amsmath,amsfonts,amssymb,graphics,graphicx,epsfig,color,times,bbm}
\usepackage{graphicx}
\newtheorem{theorem}{Theorem}
\newtheorem{lemma}{Lemma}

\begin{document}

\bibliographystyle{apsrev}

\newcommand{\Tr}{\text{Tr}}
\newcommand{\proofend}{\hfill\fbox\\\medskip }
\newcommand{\ack}{\subsection*{\normalsize \sf \textbf{Acknowledgement}}}

\title{Fine's theorem, noncontextuality, and correlations in Specker's scenario}
\author{Ravi Kunjwal}
\email{rkunj@imsc.res.in} 
\affiliation{Optics \& Quantum Information Group,
The Institute of Mathematical
Sciences, C.I.T Campus, Taramani, Chennai 600 113, India.}

\begin{abstract}
A characterization of noncontextual models which fall within the ambit of Fine's theorem is provided. In particular,
the equivalence between the existence of three notions is made explicit:
a joint probability distribution over the outcomes of all the measurements considered, a measurement-noncontextual and outcome-deterministic
(or KS-noncontextual, where `KS' stands for `Kochen-Specker') model for these measurements, and a measurement-noncontextual and factorizable 
model for them. A KS-inequality, therefore, is implied by each of these three notions. 
Following this characterization of noncontextual models that fall within the ambit of Fine's theorem,
non-factorizable noncontextual models which lie outside the domain of Fine's theorem are considered. While outcome determinism
for projective (sharp) measurements in quantum theory can be shown to follow from the assumption of preparation noncontextuality,
such a justification is not available for nonprojective (unsharp) measurements which ought to admit outcome-indeterministic response functions.
The Liang-Spekkens-Wiseman (LSW) inequality is cited as an example of a noncontextuality inequality that should hold in any noncontextual model of quantum theory 
without assuming factorizability. Three other noncontextuality
inequalities, which turn out to be equivalent to the LSW inequality under relabellings of measurement outcomes,
are derived for Specker's scenario. The polytope of correlations admissible in this scenario, given the no-disturbance condition,
is characterized.
\end{abstract}


\maketitle

\section{Introduction}

In attempts to provide a more complete description of reality than operational quantum theory in terms of a noncontextual ontological model, it is almost always assumed that 
whatever the ontic state $\lambda$ is, it must specify the outcomes of measurements exactly (an assumption called \emph{outcome determinism}) and any operational unpredictability in the measurement
outcomes is on account of coarse-graining over these ontic states $\lambda$. This paper concerns itself with what can still be said about noncontextuality
if outcome determinism is not assumed: the ontic state is not always required to fix the outcomes of measurements but only their probabilities.
The physical motivation for this becomes clear once the following questions are asked:
\begin{enumerate}
 \item Do there exist noncontextual ontological models of quantum theory where the ontic state $\lambda$ fixes the outcomes of 
measurements?\\

The Kochen-Specker theorem \cite{KS67} rules out this possibility. Let us now remove the requirement of outcome determinism, namely, that $\lambda$ fix the outcomes of 
measurements, and ask the question:
\item Do there exist noncontextual ontological models of quantum theory where the ontic state $\lambda$ fixes the \emph{probabilities}
of outcomes of measurements?\\

The Kochen-Specker theorem \cite{KS67} is silent on this question since it presumes the ontic state $\lambda$ must fix the 
outcomes of (projective) measurements. This question is most naturally addressed in the framework of generalized noncontextuality
due to Spekkens \cite{Spe05}. This is the framework adopted in this paper.
\end{enumerate}

It is well-known that, in contrast to the Kochen-Specker theorem \cite{KS67}, Bell's theorem \cite{Bell64,Bell76,Wiseman} does not require an assumption that
the ontic state $\lambda$ fixes the outcomes of the measurements. This becomes particularly clear in view of Fine's theorem \cite{Fine, Fine2} 
that, in a Bell scenario, a locally deterministic model \cite{Bell64} exists if and only if
a locally causal (or `Bell-local') model \cite{Bell76, Wiseman} exists, and how this is equivalent to requiring the existence of a joint probability distribution over
outcomes of all the measurements considered in a Bell scenario. Hence, even if the outcomes are only determined probabilistically by $\lambda$
in the local hidden variable model, Bell's theorem holds. The key issue in Bell scenarios is factorizability: the conditional independence of the outcomes of spacelike separated measurements given the ontic state
$\lambda$ of the system, 
\begin{eqnarray}
&&\xi(X_1,\dots,X_N|M_1,\dots,M_N,\lambda)\nonumber\\
&=&\xi(X_1|M_1,\lambda)\xi(X_2|M_2,\lambda)\dots\xi(X_N|M_N,\lambda),
\end{eqnarray}
where $X_i$ labels the outcome of measurement $M_i$ performed by the $i$th party, $i\in\{1,\dots,N\}$. All these response functions may be 
outcome-indeterministic, i.e., $\xi\in[0,1]$. Indeed, factorizability is a necessary consequence of any set of assumptions that may be used to derive Bell's theorem \cite{Wiseman}.

Note also that factorizability is a weaker constraint than outcome determinism since the latter implies the former but the converse
does not hold: that is, given that $\xi\in\{0,1\}$ for all the response functions above, it follows that 
$\xi(X_1,\dots,X_N|M_1,\dots,M_N,\lambda)=\delta_{X_1\dots X_N,X^{\prime}_1\dots X^{\prime}_N(\lambda)}$ and $\xi(X_i|M_i,\lambda)=\delta_{X_i,X^{\prime}_i(\lambda)}$
for all $i\in\{1\dots N\}$, where $X^{\prime}_i(\lambda)$ is the outcome assigned to measurement $M_i$ by $\lambda$. Obviously, then, 
$\xi(X_1,\dots,X_N|M_1,\dots,M_N,\lambda)=\delta_{X_1\dots X_N,X^{\prime}_1\dots X^{\prime}_N(\lambda)}=\prod_{i=1}^N\delta_{X_i,X^{\prime}_i(\lambda)}=\prod_{i=1}^N\xi(X_i|M_i,\lambda)$.
To see that the converse does not hold, it suffices to consider response functions $\xi(X_i|M_i,\lambda)\in[0,1] \quad\forall i\in\{1,\dots,N\}$ and 
define $\xi(X_1,\dots,X_N|M_1,\dots,M_N,\lambda)\equiv \prod_{i=1}^N\xi(X_i|M_i,\lambda)$, which is factorizable (by definition) but not necessarily outcome-deterministic.

On the other hand, things are not as straightforward for contextuality \cite{Spe60,Bell66,KS67}. Mathematically, both Bell-local models and 
KS-noncontextual models rely on the existence of a joint probability distribution over all measurement outcomes
in a given scenario such that this distribution reproduces the observed statistics as marginals. Given this correspondence between Bell's theorem
and the KS theorem, one may ask whether the assumption of outcome determinism is really required in the KS theorem and whether the KS theorem excludes
also all outcome-indeterministic noncontextual models
on account of Fine's theorem. This paper answers this question in the negative.

The outcome-indeterministic noncontextual models excluded by the KS theorem 
theorem are precisely the ones where factorizability holds. However, in the absence of spacelike separation between measurements one does not have a compelling
justification to assume that measurement outcomes are conditionally independent of each other given the ontic state $\lambda$. The physical meaning of factorizability
is this: that the measurement outcomes do not have any correlations that are not due to the ontic state $\lambda$ of the system. One could, on the other hand,
imagine
an adversarial situation where two measurements are correlated---which is physically possible if they are not spacelike separated---and this correlation is 
not mediated only by the ontic state $\lambda$ of the system but is perhaps encoded in the degrees of freedom of the measurement apparatus
by an adversary who wants to convince the experimenter that something nonclassical is going on (in the sense of KS-contextuality) but, really, 
it is correlated noise that's doing all the work of violating a KS inequality. The LSW inequality \cite{LSW,KG} is an example of a noncontextuality
inequality that takes this possibility into account and raises the bar for what correlations count as nonclassical. This is why
we need to consider noncontextual models which are not factorizable. Since all KS-noncontextual models are factorizable on account 
of Fine's theorem, as will become clear in Sec. III,
noncontextual models which are not factorizable are exclusively taken into account \emph{only} in the generalized definition of noncontextuality \cite{Spe05}.
This realization is a key conceptual insight of this paper, pointing to the necessity of revising the traditional analyses of KS-noncontextuality to accomodate
the generalized notion of noncontextuality \cite{Spe05}. 

Besides, just as local causality does not require the assumption of outcome determinism, a good definition of noncontextuality should also not appeal to outcome determinism (or even factorizability).
Experimental violations of Bell inequalities certify a kind of nonclassicality independent of the 
truth of quantum theory, a feature that makes Bell inequality violations an invaluable resource in device-independent protocols \cite{nsqkd}.
In contrast, a KS-noncontextual model has to refer to projective (sharp) measurements in quantum
theory and assume outcome-determinism for them in order to obtain a KS-inequality: neither of these is needed in a Bell-local model. The generalized notion of
noncontextuality offers the possibility of talking about noncontextuality without making the assumption that the operational theory is quantum theory.
The present paper, however, restricts itself to generalized noncontextuality for operational quantum theory.

The main contributions of this paper are twofold: Firstly, as noted above and proven in Sec. III, it shows the relevance of Fine's theorem in recognizing the limitations that 
the assumption of outcome determinism places on considerations of noncontextual models of quantum theory, as well as why one should worry about non-factorizable
noncontextual models. Secondly, after noting these connections between Fine's theorem, noncontextuality, and the status of outcome determinism in noncontextual models of quantum
theory \cite{odum}, a complete analysis of Specker's scenario is provided in Sec. IV for non-factorizable noncontextual models.

More specifically: In Sec. II, the notions of an operational theory and an ontological model 
of an operational theory are recalled, followed by the definition of noncontextuality due to Spekkens. Unless otherwise specified, 
`noncontextuality' will refer to this generalized notion in this paper. In particular, the traditional notion
of noncontextuality, due to Bell, Kochen, and Specker \cite{Spe60,Bell66,KS67}, will be referred to as `KS-noncontextuality'. In Sec. III, I state
Fine's theorem in the language of (generalized) noncontextuality \cite{Spe05} and discuss how this fits with the interpretation of Fine's theorem applied
to Bell scenarios. In particular, this shows why outcome-determinism is not an issue in Bell scenarios but it is an issue that 
needs to be handled with care in noncontextual models (see Ref. \cite{odum} for other compelling reasons for this). Sec. IV characterizes the correlations in Specker's 
scenario. I derive three noncontextuality inequalities
that do not assume outcome determinism or factorizability. They turn out to be equivalent to the LSW inequality under appropriate relabelling of 
measurement outcomes in this scenario. Also, the polytope of correlations admissible in Specker's scenario, given the no-disturbance condition,
is characterized by specifying all its extremal points. Section V concludes with a discussion.

\section{Operational theories and Ontological models}

\paragraph*{Operational theory.} An operational theory is specified by a triple, $(\mathcal{P},\mathcal{M},p)$, where $\mathcal{P}$ denotes 
the preparation procedures $P\in \mathcal{P}$ in the lab, $\mathcal{M}$ denotes the measurement procedures $(M,\mathcal{K}_M)\in \mathcal{M}$,
and $p:(\mathcal{P},\mathcal{M})\rightarrow [0,1]$ is the probability $p(k|P,M)$ that measurement outcome $k\in \mathcal{K}_M$
is observed when measurement procedure $(M,\mathcal{K}_M)$ is implemented following the preparation procedure $P$.

\paragraph*{Ontological model.} An ontological model $(\Lambda,\Xi,\mu)$ of an operational theory $(\mathcal{P},\mathcal{M},p)$
posits a space of ontic states $\lambda \in \Lambda$, probability densities $\mu_P: \Lambda \rightarrow [0,\infty)$
corresponding to preparation procedures $P\in\mathcal{P}$, and response functions $\xi:(\Lambda,\mathcal{M})\rightarrow [0,1]$ denoting the
probability $\xi(k|M,\lambda)$ that measurement outcome $k\in\mathcal{K}_M$ is observed when measurement procedure $(M,\mathcal{K}_M)$
is implemented and the ontic state of the system is $\lambda$. Note that $\int {\rm d}\lambda \mu_P(\lambda)=1$ and $\sum_{k\in\mathcal{K}_M}\xi(k|M,\lambda)=1$.
$\Xi$ denotes the set of response functions in the ontological model
for the measurement procedures in the operational theory. The ontological model must be empirically adequate, that is:
\begin{equation}
 p(k|P,M)=\int d\lambda \mu_P(\lambda) \xi(k|M,\lambda),
\end{equation}
for all $P\in\mathcal{P}, (M,\mathcal{K}_M)\in\mathcal{M}$. This is how an operational theory and its ontological model 
fit together.

\paragraph*{Noncontextuality.} An ontological model of an operational theory is defined to be noncontextual for prepare-and-measure experiments if it satisfies two
properties: \emph{preparation noncontextuality}, and \emph{measurement noncontextuality}. The content of preparation noncontextuality is 
captured in the following inference from the operational theory to its ontological model:

\begin{eqnarray}\nonumber
 &&p(k|P,M)=p(k|P',M), \forall k\in\mathcal{K}_M, \forall (M,\mathcal{K}_M)\in \mathcal{M}\\
 &\Rightarrow& \mu_{P}(\lambda)=\mu_{P'}(\lambda), \forall \lambda \in \Lambda.
\end{eqnarray}
That is, two preparations $P$ and $P'$ which are operationally indistinguishable are represented by identical distributions
in the ontological model. Similarly, measurement noncontextuality is simply expressed as the following inference:

\begin{eqnarray}\nonumber
 &&p(k|P,M)=p(k|P,M'),\\
 &&\forall k\in\mathcal{K}, (M,\mathcal{K}) \text{ and }(M',\mathcal{K})\in\mathcal{M}, \forall P\in \mathcal{P}\\
 &\Rightarrow& \xi(k|M,\lambda)=\xi(k|M',\lambda), \forall \lambda \in \Lambda.
\end{eqnarray}
That is, measurements $M$ and $M'$ which do not differ in their statistics relative to all preparations $P\in\mathcal{P}$ are
represented by identical response functions in the ontological model. Fine's theorem, as formulated in this paper, relates to 
the assumption of measurement noncontextuality. 

This assumption of noncontextuality is motivated by a methodological 
principle: do not introduce any differences in your explanation of two phenomena if no experiment can tell the phenomena apart. That is, if two
experimental procedures are operationally indistinguishable then they should also be indistinguishable at the ontological level, also
known as the (ontological) identity of (operational) indiscernables. For further reading,
I refer the reader to Ref.~\cite{Spe05} where this notion of noncontextuality was first defined (see also, \cite{odum}) and 
its connection with the traditional notion of KS-noncontextuality was also demonstrated.

\paragraph*{Outcome determinism} is the assumption that every response function in the ontological model is deterministic, i.e., $\xi(k|M,\lambda)\in\{0,1\}$
for all $(M,\mathcal{K}_M)\in\mathcal{M}$, $\lambda\in\Lambda$, and $\xi\in\Xi$.

Ontological models where outcome determinism doesn't hold are called outcome-indeterministic. Of the class of outcome-indeterministic
ontological models, the ones that are related to outcome-deterministic models via Fine's theorem are the models that 
satisfy \emph{factorizability}:

\paragraph*{Factorizability} is the assumption that for every jointly measurable set of measurements $\{M_s^{(S)}|s\in S\}$, the response function for every 
outcome of a joint measurement $M_S$ is the product of the response functions of measurements in the jointly measurable set: 
$\xi(k_S|M_S,\lambda)=\prod_{s\in S}\xi(k_s|M_s^{(S)},\lambda)$. Note that $k_S\in\mathcal{K}_{M_S}$ and $k_s\in\mathcal{K}_{M_s^{(S)}}$, where $\mathcal{K}_{M_S}$ is 
the Cartesian product of the outcome sets $\mathcal{K}_{M_s^{(S)}}, s\in S$.

In the next section, I will point out how these assumptions are related via Fine's theorem. This will be followed by a discussion of 
how, although factorizability is a physically motivated assumption in locally causal models, it does not admit such a motivation in the
more general case of noncontextual models. Fine's theorem thus serves to delineate a mathematical boundary between KS-noncontextual models and 
noncontextual models which are not factorizable.

\section{Fine's theorem for noncontextual models}

\begin{theorem}\label{genFine}
 Given a set of measurements $\{M_1,\dots,M_N\}$ with jointly measurable subsets $S \subset \{1,\dots,N\}$, where each measurement $M_s, s\in S$, takes 
 values labelled by $k_s\in\mathcal{K}_{M_s}$, the following propositions are equivalent:
\begin{enumerate}
\item For a given preparation $P\in\mathcal{P}$ of the system there exists a joint probability distribution $p(k_1,\dots,k_N|P)$ that recovers the marginal 
statistics for jointly measurable subsets predicted by the operational theory (such as quantum theory) under consideration, i.e., 
$\forall S \subset \{1,\dots,N\}$, $p(k_S|M_S;P)=\sum_{k_i: i \notin S} p(k_1,\dots,k_N|P)$,
where $k_S\in\mathcal{K}_{M_S}$. 

\item There exists a measurement-noncontextual and outcome-deterministic, i.e. KS-noncontextual, model for these measurements.

\item There exists a measurement-noncontextual and factorizable model for these measurements.
 \end{enumerate}
\end{theorem}

\emph{Proof.}
The proof of equivalence of the three propositions proceeds as follows: Proposition 3 $\Rightarrow$ Proposition 1, Proposition 1 $\Rightarrow$ Proposition 2, Proposition 2 $\Rightarrow$ Proposition 3.\\\\
\textbf{Proposition 3 $\Rightarrow$ Proposition 1:}\\

By Proposition 3, the assumption of measurement noncontextuality requires that the single-measurement response functions in the model be of the form $\xi(k_i|M_i;\lambda) \in [0,1]$, so
that each response function is independent of the contexts---jointly measurable subsets---that the corresponding measurement may be a part of. Of course,
the assumption of measurement noncontextuality only applies once it is verified that for any $P\in\mathcal{P}$ the operational statistics $p(k_i|M_i;P)$ of measurement $M_i$ is the same across all the jointly measurable subsets 
in which it appears. The response function is therefore conditioned only by $M_i$ and the ontic state $\lambda$ associated with the system.
Proposition 3 requires, in addition, factorizability, i.e., for all jointly measurable subsets $S \subset \{1,\dots,N\}$, $$\xi(k_S|M_S;\lambda)=\prod_{s\in S}\xi(k_s|M_s;\lambda).$$
Factorizability amounts to the assumption that the correlations between measurement outcomes are established only via the 
ontic state of the system---the measurements do not ``talk'' to each other except via $\lambda$.
Now define 
\begin{equation}
 \xi(k_1,\dots,k_N|\lambda) \equiv \prod_{i=1}^{N}\xi(k_i|M_i;\lambda),
\end{equation}
so that marginalizing this distribution over $k_i$, $i\notin S$, yields $\xi(k_S|M_S;\lambda)$ for every jointly measurable subset $S\subset \{1,\dots,N\}$.

Assuming the ontological model reproduces the operational statistics, there must exist a probability density function $\mu(\lambda|P)$ for any $P\in\mathcal{P}$, such that
\begin{equation}
 \int d\lambda \xi(k_S|M_S; \lambda)\mu(\lambda|P)=p(k_S|M_S;P).
\end{equation}
Then define
\begin{equation}
 p(k_1\dots k_N|P)\equiv \int d\lambda \xi(k_1\dots k_N|\lambda)\mu(\lambda|P),
\end{equation}
which marginalizes on $k_S$ to
\begin{eqnarray}
 p(k_S|P)&=&\sum_{k_i: i \notin S}p(k_1\dots k_N|P)\\
 &=& \int d\lambda \sum_{k_i:i \notin S}\xi(k_1\dots k_N|\lambda)\mu(\lambda|P)\\
 &=& \int d\lambda \xi(k_S|M_S;\lambda)\mu(\lambda|P)\\
 &=& p(k_S|M_S;P).
\end{eqnarray}
Thus, Proposition 3 $\Rightarrow$ Proposition 1.\\\\
\textbf{Proposition 1 $\Rightarrow$ Proposition 2:}\\

By Proposition 1, for a given $P\in\mathcal{P}$ there exists a $p(k_1\dots k_N|P)$ such that $p(k_S|M_S;P)=\sum_{k_i:i \notin S}p(k_1\dots k_N|P)$,
for all jointly measurable subsets $S \subset \{1,\dots,N\}$. Now, there exists a probability density
function $\mu(\lambda|P)$ such that 
\begin{equation}
 p(k_1\dots k_N|P)=\int d\lambda \xi(k_1\dots k_N|\lambda)\mu(\lambda|P)
\end{equation}
where $\xi(k_1\dots k_N|\lambda) \in \{0,1\}$. This is possible because any probability distribution can be decomposed as
a convex sum over deterministic distributions. Also, $p(k_j|M_j;P)=\sum_{k_i:i\neq j}p(k_1,\dots,k_N|P)$, so

\begin{equation}
 p(k_j|M_j;P)=\int d\lambda \mu(\lambda|P) \sum_{k_i:i\neq j}\xi(k_1\dots k_N|\lambda),
\end{equation}
which allows the definition
\begin{equation}
 \xi(k_j|M_j;\lambda)\equiv \sum_{k_i:i\neq j}\xi(k_1\dots k_N|\lambda) \in \{0,1\}, \forall j \in \{1\dots N\}.
\end{equation}
Since these are deterministic distributions, 
\begin{equation}
 \xi(k_1\dots k_N|\lambda)=\prod_{j=1}^{N}\xi(k_j|M_j;\lambda).
\end{equation}
Finally,
\begin{equation}
 p(k_S|M_S;P)=\int d\lambda \mu(\lambda|P)\prod_{s \in S}\xi(k_s|M_s;\lambda), 
\end{equation}
so there exists a measurement-noncontextual and outcome-deterministic model, i.e., Proposition 1 $\Rightarrow$ Proposition 2.\\
\\
\textbf{Proposition 2 $\Rightarrow$ Proposition 3:}\\\\
By Proposition 2, $\xi(k_i|M_i;\lambda) \in \{0,1\}, \forall i \in \{1\dots N\}$, such that
\begin{equation}
 p(k_S|M_S;P)=\int d\lambda \mu(\lambda|P)\prod_{s \in S}\xi(k_s|M_s;\lambda),
\end{equation}
$\forall$ jointly measurable subsets $S \subset \{1\dots N\}$. Clearly, this model is also a measurement-noncontextual 
and factorizable model because the assumption of outcome-determinism implies factorizability:
\begin{equation}
\xi(k_S|M_S;\lambda)=\prod_{s \in S}\xi(k_s|M_s;\lambda).
\end{equation}
\proofend

Note that this theorem itself is not new, but this particular reading of it in the framework of generalized noncontextuality \emph{is} new. 
In particular, the purpose of this restatement is to highlight why outcome-determinism is not an assumption
that can be taken for granted in noncontextual ontological models. Versions of this theorem have appeared in the literature
following Fine's original insight \cite{Fine,Fine2}. 
The fact that Proposition 2 implies Proposition 1 has been shown earlier in Ref.~\cite{LSW}. A similar result in the language 
of sheaf theory can be found Ref.~\cite{abramskybranden}, where the authors point out factorizability as the underlying assumption
in Bell-local and KS-noncontextual models: in effect they show the equivalence of Proposition 1 and Proposition 3. The sense in which 
Ref.~\cite{abramskybranden} refers to `non-contextuality'
is the notion of KS-noncontextuality, and while it is possible to provide a unified account of Bell-locality and KS-noncontextuality 
at a mathematical level, the generalized notion of noncontextuality \cite{Spe05} does not admit such an account. In particular,
their definition of `non-contextuality' is stronger than the Spekkens' definition of measurement noncontextuality.
Indeed, generalized noncontextuality subsumes KS-noncontextuality but is not equivalent to it.

\paragraph*{Fine's theorem for Bell scenarios.}
Translating the preceding notions from noncontextual models to Bell-local models amounts to replacing `measurement-noncontextual and outcome-deterministic' by `locally deterministic'
and `measurement-noncontextual and factorizable' by `locally causal'. 
Consider the case of two-party Bell scenarios for simplicity, although the same considerations extend to general multiparty Bell 
scenarios in a straightforward manner. A two-party Bell scenario consists of measurements $\{M_1,\dots,M_N\}$, where $\{M_1,\dots,M_n\}$, $n<N-1$, are the measurement settings 
available to one party, say Alice, and $\{M_{n+1},\dots,M_N\}$ are the measurement settings available to the other party, say Bob.
The outcomes are denoted by $k_i\in\mathcal{K}_{M_i}$ for the respective measurement settings $M_i$. The jointly measurable subsets are given 
by $S\in \{\{i,j\}|i\in\{1,\dots,n\},j\in\{n+1,\dots,N\}\}$. Bell's assumption of local causality captures the notion of
a measurement noncontextual and factorizable model:

\begin{eqnarray}\nonumber
&&p(k_S|M_S;P)\\ 
&=&p(k_i,k_j|M_i,M_j;P)\\
&=&\int {\rm d}\lambda \mu(\lambda|P)\xi(k_i|M_i,\lambda)\xi(k_j|M_j,\lambda).
\end{eqnarray}

Once factorizability is justified from Bell's assumption of local causality in this manner, Fine's theorem ensures that---so far as the existence of hidden variable models
is concerned---it is irrelevant whether the response functions for the measurement outcomes are deterministic or indeterministic.
One does not need to worry about whether outcome-determinism for measurements is justified in Bell scenarios precisely
because factorizability along with Fine's theorem absolves one of the need to provide such a justification.
The crucial point, then, is the validity of factorizability in the more general case of noncontextual models. 
In general, factorizability is not justified in noncontextual models and, following Spekkens, one must
distinguish between the issue of noncontextuality and that of outcome-determinism when considering ontological models of an 
operational theory \cite{Spe05}. If the goal is---as it should be---to obtain an experimental test of noncontextual models independent of the truth of 
quantum theory, then one needs to derive noncontextuality inequalities that do not rely on outcome-determinism at all. This is because
Fine's theorem for noncontextual models is of limited applicability---namely, outcome-indeterministic response functions which satisfy factorizability
are shown by it to achieve no more generality than is already captured by outome-deterministic response functions in a KS-noncontextual model.
Outcome-indeterministic response functions that do not satisfy factorizability are not taken into account in a KS-noncontextual model.

For ontological models of operational \emph{quantum} theory, outcome-determinism for sharp (projective) measurements can be shown to follow from 
the assumption of preparation noncontextuality \cite{Spe05}. Such a justification is not available for unsharp (nonprojective) measurements, which should therefore
be represented by outcome-indeterministic response functions. This issue has been discussed at length by Spekkens and the reader is referred
to Ref.~\cite{odum} for why and how this must be so. Therefore, to consider noncontextuality for unsharp measurements in full generality the noncontextuality
inequalities of interest are those which do not assume factorizability. An example is the LSW inequality for Specker's scenario \cite{LSW} that does not rely
on factorizability, although it does use the assumption of outcome determinism for sharp (projective) measurements.
The LSW inequality has been shown to be violated by quantum predictions \cite{KG}, thus ruling out noncontextual models of quantum theory
without invoking factorizability.
Note that the distinction between sharp and unsharp measurements is not part of the definition of a Bell-local model and one never has to worry
about this distinction to derive Bell's theorem. This distinction, however, becomes relevant for noncontextual models of quantum theory,
where the words `sharp' and `unsharp' have a clear meaning, the former referring to projective measurements and the latter to nonprojective
measurements.

In the next section, the polytope of correlations admissible in Specker's scenario is characterized.
\section{Correlations in Specker's scenario}

In this section three noncontextuality inequalities relevant to the correlations in Specker's
scenario are derived. They are shown to be equivalent to the known LSW inequality under relabelling of measurement
outcomes. This scenario involves three binary measurements, $\{M_1,M_2,M_3\}$, which are pairwise jointly measurable with outcomes 
labelled by $X_i\in\{0,1\}$ for $i\in\{1,2,3\}$. The statistics involved in Specker's scenario for a given preparation $P\in\mathcal{P}$ can be understood as a set of $12$ probabilities, $4$ for each pairwise joint measurement $M_{ij}$, 

\begin{equation}
\mathcal{S}\equiv\{p(X_iX_j|M_{ij};P)|X_i,X_j\in\{0,1\}, i,j\in\{1,2,3\},i<j\},
\end{equation}
subject to the obvious constraints of
positivity,
\begin{equation}
p(X_iX_j|M_{ij};P)\geq0 \quad \forall X_i,X_j,M_{ij},
\end{equation}
and normalization,
\begin{equation}
\sum_{X_i,X_j}p(X_iX_j|M_{ij};P)=1 \quad \forall M_{ij}.
\end{equation}

In addition to positivity and normalization, the statistics is assumed to obey the following condition:

\begin{eqnarray}\nonumber
&&\sum_{X_j}p(X_iX_j|M_{ij};P)\\
&=&\sum_{X_k}p(X_iX_k|M_{ik};P)\\
&\equiv&p(X_i|M_i;P),
\end{eqnarray}
for all $i<j,k$ where $i,j,k\in\{1,2,3\}$. Denoting $$\sum_{X_j}p(X_iX_j|M_{ij};P)\equiv p(X_i|M_i^j;P),$$ and $$\sum_{X_k}p(X_iX_k|M_{ik};P)\equiv p(X_i|M_i^k;P),$$ the condition
becomes
\begin{equation}
 p(X_i|M_i^j;P)=p(X_i|M_i^k;P)\equiv p(X_i|M_i;P).
\end{equation}

That is, the statistics of $M_i^j$, which is obtained by marginalizing the statistics of joint measurement $M_{ij}$, is identical to the statistics of $M_i^k$, which is obtained by 
marginalizing the statistics of joint measurement $M_{ik}$. If what has been measured is indeed a unique observable $M_i$ then its statistics relative to any preparation $P\in\mathcal{P}$ should
remain the same across joint measurements with different observables $M_j$ and $M_k$. Failure to meet this condition implies a failure of joint measurability: then
one can distinguish between $M_i^j$ and $M_i^k$ from their statistics relative to some preparation and they would therefore correspond to two different marginal observables $M_i^j$ and $M_i^k$ rather
than a unique observable $M_i$. This condition is often called the \emph{no-disturbance} condition in the literature on contextuality. 
Operational quantum theory obeys the no-disturbance condition for joint measurements of generalized observables (which need not be projective or sequential).

\subsection{Kochen-Specker (KS) inequalities for Specker's scenario}
The four necessary and sufficient inequalities characterizing correlations which admit a KS-noncontextual model in Specker's scenario are given by:
\begin{eqnarray}\nonumber\label{ineq1}
&&R_3\equiv p(X_1\neq X_2|M_{12},P)\\\nonumber
&+&p(X_2\neq X_3|M_{23},P)\\
&+&p(X_1\neq X_3|M_{13},P)\leq 2,
\end{eqnarray}
and
\begin{eqnarray}\nonumber\label{ineq2}
&&R_0\equiv p(X_1\neq X_2|M_{12},P)\\\nonumber
&-&p(X_2\neq X_3|M_{23},P)\\
&-&p(X_1\neq X_3|M_{13},P)\leq 0,
\end{eqnarray}
\begin{eqnarray}\nonumber\label{ineq3}
&&R_1\equiv p(X_2\neq X_3|M_{23},P)\\\nonumber
&-&p(X_1\neq X_3|M_{13},P)\\
&-&p(X_1\neq X_2|M_{12},P)\leq 0,
\end{eqnarray}
\begin{eqnarray}\nonumber\label{ineq4}
&&R_2\equiv p(X_1\neq X_3|M_{13},P)\\\nonumber
&-&p(X_1\neq X_2|M_{12},P)\\
&-&p(X_2\neq X_3|M_{23},P)\leq 0.
\end{eqnarray}

These inequalities have earlier appeared in Ref.~\cite{cabelloncycle}. A derivation is provided in Appendix \ref{deriv}. Further, these inequalities
exhibit a curious property that no two of them can be violated by the same set of experimental statistics:
\begin{lemma}\label{kslemma}
There exists no set of distributions $\{p(X_i,X_j|M_{ij},P)|(ij)\in\{(12),(23),(13)\}\}$
that can violate any two of the four KS inequalities simultaneously.
\end{lemma}
\emph{Proof.}
Denoting $w_{12}\equiv p(X_1\neq X_2|M_{12},P)$, $w_{23}\equiv p(X_2\neq X_3|M_{23},P)$, and $w_{13}\equiv p(X_1\neq X_3|M_{13},P)$, the four
KS inequalities can be rewritten as:
\begin{eqnarray}
&&R_3\equiv w_{12}+w_{23}+w_{13}\leq 2,\\
&&R_0\equiv w_{12}-w_{23}-w_{13}\leq 0,\\
&&R_1\equiv w_{23}-w_{13}-w_{12}\leq 0,\\
&&R_2\equiv w_{13}-w_{23}-w_{12}\leq 0.
\end{eqnarray}

Now, violation of each of these is equivalent to the following, since $0\leq w_{12},w_{23},w_{13}\leq 1$:
\begin{eqnarray}\nonumber
&&R_3>2\Leftrightarrow w_{12}+w_{23}+w_{13}>2,\\\nonumber
&&R_0>0\Leftrightarrow w_{12}>w_{23}+w_{13} \Rightarrow w_{12}+w_{23}+w_{13}<2,\\\nonumber
&&R_1>0\Leftrightarrow w_{23}>w_{12}+w_{13} \Rightarrow w_{12}+w_{23}+w_{13}<2\\\nonumber
&&\text{ and }w_{12}<w_{23}-w_{13},\\\nonumber
&&R_2>0\Leftrightarrow w_{13}>w_{12}+w_{23} \Rightarrow w_{12}+w_{23}+w_{13}<2\\\nonumber
&&\text{ and }w_{12}<w_{13}-w_{23}.
\end{eqnarray}
It follows that violation of each inequality above is in conflict with a violation of each of the other three inequalities. Hence, there exist 
no conceivable measurement statistics that violate any two of the four KS inequalities simultaneously.\proofend

\subsection{Noncontextuality (NC) inequalities for Specker's scenario}

Consider the \emph{predictability} of each measurement $M_k$ defined as: 
\begin{equation}
\eta_{M_k}\equiv\max_P\{2\max_{X_k}p(X_k|M_k,P)-1\}, 
\end{equation}
where $P$ is any preparation of the system. Assuming the three measurements in Specker's scenario have the same predictability $\eta_0\equiv \eta_{M_1}=\eta_{M_2}=\eta_{M_3}$,
the following noncontextuality inequalities hold:

\subsubsection{LSW inequality}
\begin{eqnarray}\nonumber\label{ncineq1}
&&R_3=p(X_1\neq X_2|M_{12},P)\\\nonumber
&+&p(X_2\neq X_3|M_{23},P)\\
&+&p(X_1\neq X_3|M_{13},P)\leq 3-\eta_0,
\end{eqnarray}
\subsubsection{Three more inequalities}
\begin{eqnarray}\nonumber\label{ncineq2}
&&R_0=p(X_1\neq X_2|M_{12},P)\\\nonumber
&-&p(X_2\neq X_3|M_{23},P)\\
&-&p(X_1\neq X_3|M_{13},P)\leq 1-\eta_0,
\end{eqnarray}
\begin{eqnarray}\nonumber\label{ncineq3}
&&R_1=p(X_2\neq X_3|M_{23},P)\\\nonumber
&-&p(X_1\neq X_3|M_{13},P)\\
&-&p(X_1\neq X_2|M_{12},P)\leq 1-\eta_0,
\end{eqnarray}
\begin{eqnarray}\nonumber\label{ncineq4}
&&R_2=p(X_1\neq X_3|M_{13},P)\\\nonumber
&-&p(X_1\neq X_2|M_{12},P)\\
&-&p(X_2\neq X_3|M_{23},P)\leq 1-\eta_0.
\end{eqnarray}
These inequalities are derived in Appendix \ref{deriv}. Note that violation of each of these inequalities implies the violation of 
the corresponding KS inequalities (recovered for $\eta_0=1$), but not conversely. 

\begin{lemma}
There exists no set of distributions $\{p(X_i,X_j|M_{ij},P)|(ij)\in\{(12),(23),(13)\}\}$
that can violate any two of the four NC inequalities simultaneously.
\end{lemma}
\emph{Proof.}
The proof trivially follows from Lemma \ref{kslemma}, since violation of any NC inequality implies violation of 
the corresponding KS inequality.\proofend

The predictability, $\eta_0$, quantifies how predictable a measurement
can be made in a variation over preparations: KS inequalities make sense only when $\eta_0=1$, i.e., it is possible to find a preparation which makes a 
given measurement perfectly predictable, a condition which is naturally satisfied by sharp (projective) measurements in quantum theory.
For the case of unsharp measurements, $\eta_0<1$, and the noncontextuality inequalities take this into account. When $\eta_0=0$, that is,
when the measurement outcomes are uniformly random (or completely unpredictable), the upper bounds in the noncontextuality inequalities become trivial
and a noncontextual model is always possible: simply ignore the system and toss a fair coin to decide whether to output $(X_i=0,X_j=1)$ or $(X_i=1,X_j=0)$ when
a pair of measurements $\{M_i,M_j\}$ is jointly implemented,
\begin{equation}
 p(X_i,X_j|M_{ij},P)=\frac{1}{2}\left(\delta_{X_i,0}\delta_{X_j,1}+\delta_{X_i,1}\delta_{X_j,0}\right).
\end{equation}
Clearly, $R_3=3$ for this, and $\eta_0=0$ since the marginal for each measurement $M_i$ is uniformly random independent of the preparation,
so the LSW inequality cannot be violated. This admits a noncontextual model since the response function for each measurement $M_i$ is a fair
coin flip independent of the system's ontic state and also of which other measurement it is jointly implemented with. The key feature that the
LSW inequality captures is this: that it is not possible to have a high degree of anticorrelation $R_3$ and a high degree of predictability
$\eta_0$ in a noncontextual model, and that there is a tradeoff between the two, given here by $R_3+\eta_0\leq 3$. Contextuality in this sense signifies
the ability to generate (anti)correlations which violate this tradeoff for values of $\eta_0<1$: the case $\eta_0=1$, as mentioned, is already covered
by the usual KS inequalities, and for $\eta_0=0$ there is no nontrivial tradeoff imposed by noncontextual models.

\subsection{Equivalence under relabelling of measurement outcomes}
The four NC inequalities (also the KS inequalities) are equivalent under relabelling measurement outcomes: To go from $R_3\leq 3-\eta_0$ to 
$R_0\leq 1-\eta_0$, simply relabel the measurement outcomes of $M_3$ as $X_3\rightarrow X'_3=1-X_3$, so that after the relabelling (denoted by primed quantities):
$w'_{12}=w_{12},w'_{23}=1-w_{23},w'_{13}=1-w_{13}$, and
$R'_3\equiv w'_{12}+w'_{23}+w'_{13}\leq 3-\eta_0$ becomes $w_{12}+(1-w_{23})+(1-w_{13})\leq 3-\eta_0$
which can be rewritten as $R'_3=R_0=w_{12}-w_{23}-w_{13}\leq 1-\eta_0$. Similarly, relabelling measurement outcomes of $M_2$ takes $R_3\leq 3-\eta_0$
to $R_2\leq 1-\eta_0$ and relabelling measurement outcomes of $M_1$ takes $R_3\leq 3-\eta_0$ to $R_1\leq 1-\eta_0$.

\subsection{Quantum violation of noncontextuality inequalities for Specker's scenario}
Quantum realization of Specker's scenario involves three unsharp qubit POVMs $M_k=\{E^k_0,E^k_1\}, k\in\{1,2,3\}$, where the effects are given by:
\begin{equation}
 E^k_{X_k}\equiv\frac{1}{2}I+(-1)^{X_k}\frac{\eta}{2}\vec{\sigma}.\hat{n}_k,\quad X_k\in\{0,1\}, 0\leq\eta\leq1.
\end{equation}
These can be rewritten as:
\begin{equation}
E^k_{X_k}=\eta\Pi^k_{X_k}+(1-\eta)\frac{I}{2}, 
\end{equation}
where $\Pi^k_{X_k}=\frac{1}{2}(I+(-1)^{X_k}\vec{\sigma}.\hat{n}_k)$ are the corresponding projectors. That is, $M_k$ is a noisy version of the projective measurement
of spin along the $\hat{n}_k$ direction, where the sharpness of the POVM is given by $\eta$. In this case, $p(X_k|M_k,P)=\Tr(\rho_P E^k_{X_k})$, where $\rho_P$ is the density matrix for preparation $P$ of the system and the predictability can be easily shown to be $\eta$: the preparation 
maximizing $\eta_{M_k}$ is a pure state along the $\hat{n}_k$ axis, i.e., $\rho_P=\Pi^k_{X_k}$.

Quantum violation of the LSW inequality has already been shown in Ref. \cite{KG}. On account of the equivalence of the four NC inequalities
under relabelling of measurement outcomes, the violation of the other three NC inequalities besides LSW follows from appropriate relabellings of
measurement outcomes in the quantum violation demonstrated in Ref. \cite{KG}.

\subsection{Specker polytope}
The statistics allowed in Specker's scenario, given that the no-disturbance condition holds, can be understood as a convex polytope in $\mathbb{R}^6$
with $12$ extreme points or vertices,
$8$ of which are deterministic and $4$ indeterministic. The measurement statistics are given by the vector of $12$ probabilities 
$\vec{v}(P)=(v^{ij}_{X_iX_j}(P)|X_i,X_j\in\{0,1\}, i,j\in\{1,2,3\},i<j)$, where $v^{ij}_{X_iX_j}(P)\equiv p(X_iX_j|M_{ij};P)$, constrained by the positivity, normalization 
and no-disturbance conditions which reduce the number of independent probabilities in $\vec{v}(P)$ from $12$ to $6$.

The deterministic vertices, which admit KS-noncontextual models, correspond to the $8$ possible tripartite joint distributions of the form,
$p(X_1,X_2,X_3|P)\equiv \delta_{X_1,X_1(P)}\delta_{X_2,X_2(P)},\delta_{X_3,X_3(P)}$, where $X_1(P),X_2(P),X_3(P)\in \{0,1\}$. The deterministic vertex $\vec{v}(P)$
can be obtained from this joint distribution as $v^{ij}_{X_iX_j}(P)=\sum_{X_k, k\neq i,j}p(X_1,X_2,X_3|P)=\delta_{X_i,X_i(P)}\delta_{X_j,X_j(P)}$.
These vertices are labelled lexicographically, $(X_1(P),X_2(P),X_3(P))$ as the decimal equivalent of binary number $X_1(P)X_2(P)X_3(P)$:

\begin{align}
\vec{v}_0(P): v^{12}_{00}(P)=v^{23}_{00}(P)=v^{13}_{00}(P)=1,\\
\vec{v}_1(P): v^{12}_{00}(P)=v^{23}_{01}(P)=v^{13}_{01}(P)=1,\\
\vec{v}_2(P): v^{12}_{01}(P)=v^{23}_{10}(P)=v^{13}_{00}(P)=1,\\
\vec{v}_3(P): v^{12}_{01}(P)=v^{23}_{11}(P)=v^{13}_{01}(P)=1,\\
\vec{v}_4(P): v^{12}_{10}(P)=v^{23}_{00}(P)=v^{13}_{10}(P)=1,\\
\vec{v}_5(P): v^{12}_{10}(P)=v^{23}_{01}(P)=v^{13}_{11}(P)=1,\\
\vec{v}_6(P): v^{12}_{11}(P)=v^{23}_{10}(P)=v^{13}_{10}(P)=1,\\
\vec{v}_7(P): v^{12}_{11}(P)=v^{23}_{11}(P)=v^{13}_{11}(P)=1.
\end{align}
Note that these deterministic vertices satisfy all the four KS inequalities, Eqs. (\ref{ineq1})-(\ref{ineq4}), and therefore also the four noncontextuality 
inequalities, Eqs. (\ref{ncineq1})-(\ref{ncineq4}). That is, they admit a KS-noncontextual model. Indeed, the convex set that these $8$ extreme points 
define is a KS-noncontextuality polytope, analogous to a Bell polytope in a Bell scenario. This polytope is a 
subset of the larger Specker polytope which in addition to these $8$ vertices includes the $4$ indeterministic vertices in Specker's scenario.

The indeterministic vertices, which do \emph{not} admit KS-noncontextual models, correspond to the $4$ sets of pairwise joint distributions given by:
\begin{eqnarray}
&&\vec{v}_8(P):\\
&&v^{12}_{01}(P)=v^{12}_{10}(P)=\frac{1}{2},\\
&&v^{23}_{00}(P)=v^{23}_{11}(P)=\frac{1}{2},\\
&&v^{13}_{00}(P)=v^{13}_{11}(P)=\frac{1}{2},
\end{eqnarray}

\begin{eqnarray}
&&\vec{v}_9(P):\\
&&v^{12}_{00}(P)=v^{12}_{11}(P)=\frac{1}{2},\\
&&v^{23}_{01}(P)=v^{23}_{10}(P)=\frac{1}{2},\\
&&v^{13}_{00}(P)=v^{13}_{11}(P)=\frac{1}{2},
\end{eqnarray}

\begin{eqnarray}
&&\vec{v}_{10}(P):\\
&&v^{12}_{00}(P)=v^{12}_{11}(P)=\frac{1}{2},\\
&&v^{23}_{00}(P)=v^{23}_{11}(P)=\frac{1}{2},\\
&&v^{13}_{01}(P)=v^{13}_{10}(P)=\frac{1}{2},
\end{eqnarray}

\begin{eqnarray}
&&\vec{v}_{11}(P):\\
&&v^{12}_{01}(P)=v^{12}_{10}(P)=\frac{1}{2},\\
&&v^{23}_{01}(P)=v^{23}_{10}(P)=\frac{1}{2},\\
&&v^{13}_{01}(P)=v^{13}_{10}(P)=\frac{1}{2}.
\end{eqnarray}
The vertex $\vec{v}_8(P)$ violates inequalities (\ref{ineq2}) and (\ref{ncineq2}) ($\eta_0>0$), $\vec{v}_9(P)$ violates inequalities (\ref{ineq3})
and (\ref{ncineq3}) ($\eta_0>0$), $\vec{v}_{10}(P)$ violates inequalities (\ref{ineq4}) and (\ref{ncineq4}) ($\eta_0>0$),
and $\vec{v}_{11}(P)$ violates inequalities (\ref{ineq1}) and (\ref{ncineq1}) ($\eta_0>0$). 
Note that these vertices are equivalent under relabellings, that is, $\vec{v}_8(P)$ turns to $\vec{v}_{11}(P)$ on relabelling outcomes
of $M_3$, $\vec{v}_9(P)$ to $\vec{v}_{11}(P)$ on relabelling outcomes of $M_1$, and 
$\vec{v}_{10}(P)$ to $\vec{v}_{11}(P)$ on relabelling outcomes of $M_2$. Note that the vertex $\vec{v}_{11}(P)$ corresponds to the `overprotective seer' (OS)
correlations of Ref. \cite{LSW} which maximally violate the LSW inequality when $\eta_0<1$.

\subsection{Limitations of the joint probability distribution criterion for deciding contextuality}
All the Bell-Kochen-Specker type analyses of contextuality ultimately hinge on ruling out the existence of a joint probability
distribution that reproduces the operational statistics of various jointly measurable observables as marginals. Deciding whether
such a joint distribution exists is called a marginal problem \cite{chavesfritz}. That this is a limited criterion to 
decide the question of contextuality without also making the assumption of outcome determinism or factorizability is borne out by correlations 
in Specker's scenario that lie outside the polytope of correlations admissible in KS-noncontextual models but are realizable in noncontextual models.
Violation of the LSW inequality by unsharp measurements in quantum theory rules out such noncontextual models \cite{KG}. 

Once outcome determinism for unsharp measurements (ODUM, cf.\cite{odum}) is abandoned,
the existence of a joint distribution is no longer necessary to characterize noncontextual models.
Further, in the case of an arbitrary operational theory which isn't quantum theory it isn't obvious whether outcome-determinism for 
measurements can at all be justified from the assumption of preparation and measurement noncontextuality. An experimentally interesting
and robust noncontextuality inequality should not assume that the operational theory describing the experiment is quantum theory and 
instead derive from the assumption of noncontextuality alone, given some operational equivalences between preparation procedures or 
measurement procedures. Violation of the LSW inequality only indicates that \emph{quantum theory} does not admit a noncontextual ontological model.
The ideal to aspire for is something akin to Bell inequalities which are theory-independent. That such an ideal is achievable will be shown 
in a forthcoming paper \cite{kunjspek}.

\section{Conclusion}
To summarize, the chief takeaways from this paper are the following:
\begin{enumerate}
 \item Fine's theorem for noncontextual models only applies in cases where the correlations between measurement outcomes are mediated exclusively by 
the ontic state $\lambda$ of the system. When this is not the case and factorizability fails, it's possible that the measurement outcomes share correlations
that are not on account of the measured system but an artifact of the measurement apparatus. Considering noncontextual models which are not factorizable allows
one to handle this situation.
 \item The no-disturbance polytope of Specker's scenario admits 4 indeterministic extremal points, related to each other by relabellings of measurement outcomes, that
are related to the `OS box' of Ref. \cite{LSW}. Corresponding to these 4 extremal points are 4 Kochen-Specker inequalities assuming outcome determinism, and
4 noncontextuality inequalities that do not assume outcome determinism.
\end{enumerate}
All this points out the need to further investigate how a failure of outcome determinism or factorizability in the case of more well-known KS inequalities should be handled.
Another open question is how to derive noncontextuality inequalities for arbitrary operational theories, rather than just quantum theory,
without any assumption of outcome determinism or factorizability. These questions will be taken up in future work.
\section*{Acknowledgments}
I would like to thank Matt Pusey, Rob Spekkens and Tobias Fritz for useful discussions and the Perimeter Institute, where part of this work was carried out, for
hospitality. This work was made possible in part through the support of a grant from the John Templeton Foundation.

\begin{appendix}
\section{Constraints on the operational statistics from normalization and no-disturbance}\label{opconstraints}
The notation here is simplified as follows: the measurements are denoted by $e\equiv M_1, f\equiv M_2, g\equiv M_3$, and their outcomes 
by $e_k\equiv (X_1=k)$, $f_k\equiv (X_2=k)$, $g_k\equiv (X_3=k)$, where $k\in\{0,1\}$. Thus there are three binary observables, $e, f, g$,
each taking values in $\{0,1\}$ and measured on a system prepared according to some preparation $P$. $e_0$ denotes the outcome $e=0$ and $e_1$ denotes $e=1$. 
Analogous notation applies for outcomes of $f$ and $g$ as well. The probability distributions on these observables associated with 
the preparation $P$ are denoted by $w_P(e)\equiv\{w_P(e_0),w_P(e_1)\},w_P(f)\equiv\{w_P(f_0),w_P(f_1)\},w_P(g)\equiv\{w_P(g_0),w_P(g_1)\}$.
The experimental statistics correspond to the joint measurement of every pair of observables: 
\begin{eqnarray}\nonumber
w_P(e,f)&\equiv&\{w_P(e_0,f_0),w_P(e_0,f_1),w_P(e_1,f_0),w_P(e_1,f_1)\},\\\nonumber
w_P(f,g)&\equiv&\{w_P(f_0,g_0),w_P(f_0,g_1),w_P(f_1,g_0),w_P(f_1,g_1)\},\\\nonumber
w_P(e,g)&\equiv&\{w_P(e_0,g_0),w_P(e_0,g_1),w_P(e_1,g_0),w_P(e_1,g_1)\}.
\end{eqnarray}

In addition to the usual positivity and normalization constraints for probability distributions,
the no-disturbance condition on the pairwise joint distributions yields:
\begin{eqnarray}\nonumber
w_P(e_0,f_0)+w_P(e_0,f_1)&=&w_P(e_0,g_0)+w_P(e_0,g_1)\\\nonumber
&\equiv& w_P(e_0),\\\nonumber
\Rightarrow w_P(e_1,f_0)+w_P(e_1,f_1)&=&w_P(e_1,g_0)+w_P(e_1,g_1)\\\nonumber
&\equiv& w_P(e_1),\\\nonumber
w_P(f_0,g_0)+w_P(f_0,g_1)&=&w_P(e_0,f_0)+w_P(e_1,f_0)\\\nonumber
&\equiv& w_P(f_0),\\\nonumber
\Rightarrow w_P(f_1,g_0)+w_P(f_1,g_1)&=&w_P(e_0,f_1)+w_P(e_1,f_1)\\\nonumber
&\equiv& w_P(f_1),\\\nonumber
w_P(e_0,g_0)+w_P(e_1,g_0)&=&w_P(f_0,g_0)+w_P(f_1,g_0)\\\nonumber
&\equiv& w_P(g_0),\\\nonumber
\Rightarrow w_P(e_0,g_1)+w_P(e_1,g_1)&=&w_P(f_0,g_1)+w_P(f_1,g_1)\\\nonumber
&\equiv& w_P(g_1).
\end{eqnarray}
Normalization gets rid of three parameters out of the twelve in the experimental statistics while no-disturbance eliminates
three more parameters. There are, therefore, six independent parameters describing the experimental statistics:
\begin{eqnarray}
 w_{12}=w_P(e_0,f_1)+w_P(e_1,f_0),\\
 w_{23}=w_P(f_0,g_1)+w_P(f_1,g_0),\\
 w_{13}=w_P(e_0,g_1)+w_P(e_1,g_0),\\
 p_1\equiv w_P(e_0),\\
 p_2\equiv w_P(f_0),\\
 p_3\equiv w_P(g_0),
\end{eqnarray}
subject to $0\leq w_{12},w_{23},w_{13},p_1,p_2,p_3\leq 1$.
Using the no-disturbance and normalization conditions:
\begin{eqnarray}\nonumber
w_P(e_0,f_1)=\frac{w_{12}+p_1-p_2}{2}&,& w_P(e_1,f_0)=\frac{w_{12}-p_1+p_2}{2},\\\nonumber
w_P(e_0,f_0)=\frac{p_1+p_2-w_{12}}{2}&,& w_P(e_1,f_1)=1-\frac{w_{12}+p_1+p_2}{2},\\\nonumber
w_P(f_0,g_1)=\frac{w_{23}+p_2-p_3}{2}&,& w_P(f_1,g_0)=\frac{w_{23}-p_2+p_3}{2},\\\nonumber
w_P(f_0,g_0)=\frac{p_2+p_3-w_{23}}{2}&,& w_P(f_1,g_1)=1-\frac{w_{23}+p_2+p_3}{2},\\\nonumber
w_P(e_0,g_1)=\frac{w_{13}+p_1-p_3}{2}&,& w_P(e_1,g_0)=\frac{w_{13}-p_1+p_3}{2},\\\nonumber
w_P(e_0,g_0)=\frac{p_1+p_3-w_{13}}{2}&,& w_P(e_1,g_1)=1-\frac{w_{13}+p_1+p_3}{2}.
\end{eqnarray}
The positivity requirements on these translate to the following inequalities:
\begin{eqnarray}
 |p_1-p_2|&\leq& w_{12}\leq p_1+p_2\leq 2-w_{12},\\
 |p_2-p_3|&\leq& w_{23}\leq p_2+p_3\leq 2-w_{23},\\
 |p_1-p_3|&\leq& w_{13}\leq p_1+p_3\leq 2-w_{13}.
\end{eqnarray}

\section{Deriving the KS and NC inequalities}\label{deriv}

\subsection{KS inequalities}
The KS inequalities derive from the existence of a joint probability distribution $p(X_1X_2X_3)$ such that
$p(X_iX_j|M_{ij},P)=\sum_{X_k}p(X_1X_2X_3)$, where $i,j,k$ are distinct indices in $\{1,2,3\}$. Therefore
the following must hold:
\begin{eqnarray}
p(001)&=&p(00|M_{12},P)-p(000),\nonumber\\
p(010)&=&p(00|M_{13},P)-p(000),\nonumber\\
p(100)&=&p(00|M_{23},P)-p(000),\nonumber\\
p(011)&=&p(01|M_{12},P)-p(010)\nonumber\\
&=&p(01|M_{12},P)-p(00|M_{13},P)+p(000),\nonumber\\
p(101)&=&p(10|M_{12},P)-p(100)\nonumber\\
&=&p(10|M_{12},P)-p(00|M_{23},P)+p(000),\nonumber\\
p(110)&=&p(10|M_{13},P)-p(100)\nonumber\\
&=&p(10|M_{13},P)-p(00|M_{23},P)+p(000),\nonumber\\
p(111)&=&1-p(00|M_{12},P)-p(01|M_{12},P)-p(10|M_{12},P)\nonumber\\
&-&p(10|M_{13},P)+p(00|M_{23},P)-p(000).\nonumber
\end{eqnarray}
Expressing the probabilities in terms of the six free parameters identified earlier, namely, 
the anticorrelation probabilities, $w_{12},w_{23},w_{13}$, and the marginals $p_1,p_2,p_3$, the positivity constraints, $0\leq p(X_1X_2X_3)\leq1$, require:
\begin{eqnarray}
&&0\leq p(000)\leq 1,\nonumber\\
&&0\leq p(001)\leq 1\nonumber\\
&&\Leftrightarrow p_1+p_2-2\leq w_{12}\leq p_1+p_2-2p(000)\nonumber\\
&&0\leq p(010)\leq 1\nonumber\\
&&\Leftrightarrow p_1+p_3-2\leq w_{13}\leq p_1+p_3-2p(000)\nonumber\\
&&0\leq p(100)\leq 1\nonumber\\
&&\Leftrightarrow p_2+p_3-2\leq w_{23}\leq p_2+p_3-2p(000)\nonumber\\
&&0\leq p(011)\leq 1\nonumber\\
&&\Leftrightarrow p_2+p_3-2p(000)\leq w_{12}+w_{13}\leq 2-2p(000)+p_2+p_3\nonumber\\
&&0\leq p(101)\leq 1\nonumber\\
&&\Leftrightarrow p_1+p_3-2p(000)\leq w_{12}+w_{23}\leq 2-2p(000)+p_1+p_3\nonumber\\
&&0\leq p(110)\leq 1\nonumber\\
&&\Leftrightarrow p_1+p_2-2p(000)\leq w_{13}+w_{23}\leq 2-2p(000)+p_1+p_2\nonumber\\
&&0\leq p(111)\leq 1\nonumber\\
&&\Leftrightarrow -2p(000)\leq w_{12}+w_{23}+w_{13}\leq 2-2p(000).\nonumber
\end{eqnarray}
Combining the inequalities to eliminate $p(000)$, and using the fact that $0\leq p(000)\leq1$:
\begin{eqnarray}
&&0\leq p(000)\leq1, 0\leq p(111)\leq 1\nonumber\\
&\Rightarrow&-2p(000)\leq 0\leq w_{12}+w_{23}+w_{13}\leq 2-2p(000)\leq 2,\nonumber\\
&&0\leq p(010)\leq1, 0\leq p(101)\leq1\nonumber\\
&\Rightarrow&0\leq w_{12}+w_{23}-w_{13}\leq 2 \leq 4-2p(000),\nonumber\\
&&0\leq p(110)\leq1, 0\leq p(001)\leq1\nonumber\\
&\Rightarrow&0\leq w_{23}+w_{13}-w_{12}\leq 2 \leq 4-2p(000),\nonumber\\
&&0\leq p(011)\leq1, 0\leq p(100)\leq1\nonumber\\
&\Rightarrow&0\leq w_{12}+w_{13}-w_{23}\leq 2 \leq 4-2p(000).\nonumber
\end{eqnarray}
Of these, the KS inequalities, which are not trivially true by normalization and positivity,
are the following:
\begin{eqnarray}
&&R_3\equiv w_{12}+w_{23}+w_{13}\leq 2,\\
&&R_0\equiv w_{12}-w_{23}-w_{13}\leq 0,\\
&&R_1\equiv w_{23}-w_{12}-w_{13}\leq 0,\\
&&R_2\equiv w_{13}-w_{12}-w_{23}\leq 0.
\end{eqnarray}
Note that $p(000)\leq p_1,p_2,p_3$, since $p_1=p(000)+p(001)+p(010)+p(011)$, etc. As long as the KS inequalities 
are satisfied, one can define a joint probability distribution by choosing a suitable 
$p(000)\leq\min\{p_1,p_2,p_3\}$.

To summarize, there are following constraints on the six parameters, $\{w_{12},w_{23},w_{13},p_1,p_2,p_3\}$, characterizing the
polytope of KS-noncontextual correlations:
\begin{eqnarray}
&&0\leq p_1,p_2,p_3,w_{12},w_{23},w_{13}\leq 1,\\
&&|p_1-p_2|\leq w_{12}\leq \min\{p_1+p_2, 2-p_1-p_2\},\\
&&|p_2-p_3|\leq w_{23}\leq \min\{p_2+p_3, 2-p_2-p_3\},\\
&&|p_1-p_3|\leq w_{13}\leq \min\{p_1+p_3, 2-p_1-p_3\},\\
&&w_{12}+w_{23}+w_{13}\leq 2,\\
&&w_{12}-w_{23}-w_{13}\leq 0,\\
&&w_{23}-w_{12}-w_{13}\leq 0,\\
&&w_{13}-w_{12}-w_{23}\leq 0.
\end{eqnarray}

\subsection{NC inequalities}
In deriving the NC inequalities, I closely follow the derivation of the LSW inequality in Ref.~\cite{LSW}. For a more detailed explication of the 
principles underlying this derivation, the reader may consult Ref.~\cite{odum}. The assumptions used are: measurement noncontextuality
and preparation noncontextuality on account of the fact that in operational quantum theory, preparation noncontextuality implies outcome determinism
for sharp measurements \cite{Spe05}.

Define
\begin{eqnarray}
&&R_3(\lambda)\equiv w_{12}(\lambda)+w_{23}(\lambda)+w_{13}(\lambda),\\
&&R_0(\lambda)\equiv w_{12}(\lambda)-w_{23}(\lambda)-w_{13}(\lambda),\\
&&R_1(\lambda)\equiv w_{23}(\lambda)-w_{12}(\lambda)-w_{13}(\lambda),\\
&&R_2(\lambda)\equiv w_{13}(\lambda)-w_{12}(\lambda)-w_{23}(\lambda),
\end{eqnarray}
where
$w_{ij}(\lambda)\equiv \xi(X_i\neq X_j|M_{ij};\lambda)$, for all $(ij)\in\{(12),(23),(13)\}$. Note that for any given preparation, the ontological model
associates a positive density $\mu(\lambda|P)\geq0$, where $\int{\rm d}\lambda\mu(\lambda|P)=1$, and $p(X|M,P)=\int{\rm d}\lambda\mu(\lambda|P)\xi(X|M;\lambda)$,
where $\xi(X|M;\lambda)\in[0,1]$ is the response function of outcome $X$ when measurement $M$ is performed and the system's ontic state is $\lambda$.
Therefore:
\begin{eqnarray}
R_3=\int {\rm d}\lambda \mu(\lambda|P)R_3(\lambda)\leq\max_{\lambda}R_3(\lambda),\\
R_0=\int {\rm d}\lambda \mu(\lambda|P)R_0(\lambda)\leq\max_{\lambda}R_0(\lambda),\\
R_1=\int {\rm d}\lambda \mu(\lambda|P)R_1(\lambda)\leq\max_{\lambda}R_1(\lambda),\\
R_2=\int {\rm d}\lambda \mu(\lambda|P)R_2(\lambda)\leq\max_{\lambda}R_2(\lambda).
\end{eqnarray}

To maximize $R_3(\lambda)$ in this noncontextual model, one needs to maximize each anticorrelation term $w_{12}(\lambda),w_{23}(\lambda),w_{13}(\lambda)$.
To maximize $R_0(\lambda)$, maximize $w_{12}(\lambda)$ and minimize $w_{23}(\lambda), w_{13}(\lambda)$. Similarly,
to maximize $R_1(\lambda)$, maximize $w_{23}(\lambda)$ and minimize $w_{12}(\lambda), w_{13}(\lambda)$, and 
to maximize $R_2(\lambda)$, maximize $w_{13}(\lambda)$ and minimize $w_{12}(\lambda), w_{23}(\lambda)$.

The single measurement response functions are given by
\begin{equation}
 \xi(X_i|M_i;\lambda)=\eta\delta_{X_i,X_i(\lambda)}+(1-\eta)\left(\frac{1}{2}\delta_{X_i,0}+\frac{1}{2}\delta_{X_i,1}\right),
\end{equation}
$i\in\{1,2,3\}$, in keeping with the assumption of outcome determinism for projectors but not so for nonprojective positive operators \cite{odum}. 
The general form the pairwise response function for measurements $\{M_i,M_j\}$ is given by:
\begin{eqnarray}
\xi(X_i,X_j|M_{ij};\lambda)&=&\alpha \delta_{X_i,X_i(\lambda)}\delta_{X_j,X_j(\lambda)}\\
&+&\beta \delta_{X_i,X_i(\lambda)}\left(\frac{1}{2}\delta_{X_j,0}+\frac{1}{2}\delta_{X_j,1}\right)\nonumber\\
&+&\gamma \left(\frac{1}{2}\delta_{X_i,0}+\frac{1}{2}\delta_{X_i,1}\right)\delta_{X_j,X_j(\lambda)}\nonumber\\
&+&\delta \left(\frac{1}{2}\delta_{X_i,0}\delta_{X_j,0}+\frac{1}{2}\delta_{X_i,1}\delta_{X_j,1}\right)\nonumber\\
&+&\epsilon \left(\frac{1}{2}\delta_{X_i,0}\delta_{X_j,1}+\frac{1}{2}\delta_{X_i,1}\delta_{X_j,0}\right).\nonumber
\end{eqnarray}
The marginals are
\begin{eqnarray}
\xi(X_i|M_{ij};\lambda)&=&(\alpha+\beta)\delta_{X_i,X_i(\lambda)}\\
&+&(\gamma+\delta+\epsilon)\left(\frac{1}{2}\delta_{X_i,0}+\frac{1}{2}\delta_{X_i,1}\right),\nonumber
\end{eqnarray}
and
\begin{eqnarray}
\xi(X_j|M_{ij};\lambda)&=&(\alpha+\gamma)\delta_{X_j,X_j(\lambda)}\nonumber\\
&+&(\beta+\delta+\epsilon)\left(\frac{1}{2}\delta_{X_j,0}+\frac{1}{2}\delta_{X_j,1}\right),
\end{eqnarray}
so that the following must hold on account of $\xi(X_i|M_{ij};\lambda)=\xi(X_i|M_i;\lambda)$ and $\xi(X_j|M_{ij};\lambda)=\xi(X_j|M_j;\lambda)$:
\begin{equation}
\alpha+\beta=\alpha+\gamma=\eta,
\end{equation}
\begin{equation}
\gamma+\delta+\epsilon=\beta+\delta+\epsilon=1-\eta.
\end{equation}

To maximize anticorrelation $w_{ij}(\lambda)$: the $\beta$ and $\gamma$ terms yield correlation as often as anticorrelation, 
so $\beta=\gamma=0$. The $\delta$ term always yields correlation, so $\delta=0$. Only $\alpha$ and $\epsilon$ terms 
allow for more anticorrelation than correlation. This means $\alpha=\eta$ and $\epsilon=1-\eta$. The pairwise response function
maximizing anticorrelation $w_{ij}(\lambda)$ is given by
\begin{eqnarray}
&&\xi(X_iX_j|M_{ij};\lambda)\nonumber\\
&=&\eta\delta_{X_i,X_i(\lambda)}\delta_{X_j,X_j(\lambda)}\nonumber\\
&+&(1-\eta)\left(\frac{1}{2}\delta_{X_i,0}\delta_{X_j,1}+\frac{1}{2}\delta_{X_i,1}\delta_{X_j,0}\right).\nonumber
\end{eqnarray}
This maximizing response function constrains the anticorrelation probability as:
\begin{equation}
1-\eta\leq w_{ij}(\lambda)\leq 1.
\end{equation}

To minimize anticorrelation $w_{ij}(\lambda)$: the $\beta$ and $\gamma$ terms yield correlation as often as anticorrelation, 
so $\beta=\gamma=0$. The $\epsilon$ term always yields anticorrelation, so $\epsilon=0$. Only $\alpha$ and $\delta$ terms 
allow for more correlation than anticorrelation. This means $\alpha=\eta$ and $\delta=1-\eta$. The pairwise response function
minimizing anticorrelation $w_{ij}(\lambda)$ is given by
\begin{eqnarray}
&&\xi(X_iX_j|M_{ij};\lambda)\nonumber\\
&=&\eta\delta_{X_i,X_i(\lambda)}\delta_{X_j,X_j(\lambda)}\nonumber\\
&+&(1-\eta)\left(\frac{1}{2}\delta_{X_i,0}\delta_{X_j,0}+\frac{1}{2}\delta_{X_i,1}\delta_{X_j,1}\right).\nonumber
\end{eqnarray}
This minimizing response function constrains the anticorrelation probability as:
\begin{equation}
0\leq w_{ij}(\lambda)\leq \eta.
\end{equation}

$R_3(\lambda)$ is maximized by considering the response function maximizing anticorrelation for each of $w_{ij}(\lambda)$,
and noting that of the eight possible assignments $\lambda\rightarrow (X_1(\lambda),X_2(\lambda),X_3(\lambda)) \in \{0,1\}^3$,
the assignments maximizing $R_3(\lambda)$ are $\{(001),(010),(011),(100),(101),(110)\}$, each of which has two anticorrelated pairs
and a third correlated pair such that the anticorrelation probability becomes
$$\max_{\lambda}R_3(\lambda)=2\eta+3(1-\eta)=3-\eta,$$ and therefore
\begin{equation}
R_3\leq 3-\eta,
\end{equation}
the LSW inequality.

$R_0(\lambda)$ is maximized by considering the response function maximizing anticorrelation for $w_{12}(\lambda)$ and
response functions minimizing anticorrelation for $w_{23}(\lambda)$ and $w_{13}(\lambda)$. Noting that of the eight possible
assignments $\lambda\rightarrow (X_1(\lambda),X_2(\lambda),X_3(\lambda)) \in \{0,1\}^3$,
the assignments maximizing $R_0(\lambda)$ are $\{(010),(011),(100),(101)\}$:
for $\{(010),(101)\}$, $w_{12}(\lambda)=1$, $w_{23}(\lambda)=\eta$, and $w_{13}(\lambda)=0$, and 
for $\{(011),(100)\}$, $w_{12}(\lambda)=1$, $w_{23}(\lambda)=0$, and $w_{13}(\lambda)=\eta$, so that
$$\max_{\lambda}R_0(\lambda)=1-\eta,$$ and therefore
\begin{equation}
R_0\leq 1-\eta.
\end{equation}
Similarly, the NC inequalities for $R_1$ and $R_2$ follow: $R_1\leq 1-\eta$ and $R_2\leq 1-\eta$.
\end{appendix}


\begin{thebibliography}{}

\bibitem{KS67} S.~Kochen and E.~P.~Specker, J.~Math. \& Mech. {\bf 17}, 59 (1967).

\bibitem{Spe60} E. P. Specker, Dialectica {\bf 14}, 239-246
    (1960); English translation: 
    M. P. Seevinck, arXiv preprint arXiv:1103.4537v3 (2011).

\bibitem{Bell66} J.S. Bell, Rev. Mod. Phys. 38, 447 (1966).

\bibitem{Spe05} R.\ W.\ Spekkens, Phys. Rev. A {\bf 71}, 052108 (2005).

\bibitem{Bell64} J.~S. Bell, Physics, {\bf 1}, 195 (1964).

\bibitem{Bell76} J.~S. Bell, Epistemological Lett. 9 (1976) (Reproduced in Bell M, Gottfried K and Veltman M (ed) 2001 John S Bell on the Foundations of Quantum Mechanics (Singapore: World Scientific)).

\bibitem{Wiseman} H.~M. Wiseman, J. Phys. A: Math. Theor. 47 424001 (2014).

\bibitem{Fine} A. Fine, Phys. Rev. Lett. {\bf 48}, 291 (1982).

\bibitem{Fine2} A. Fine, J. Math. Phys. 23, 1306 (1982).

\bibitem{nsqkd} J. Barrett, L. Hardy, A. Kent, Phys. Rev. Lett. 95, 010503 (2005).

\bibitem{odum} R.W. Spekkens, Found. Phys. {\bf 44}, 11, 1125 (2014).

\bibitem{LSW} Y.C. Liang, and R.W. Spekkens, and H.M. Wiseman, Phys. Rep. {\bf 506}, 1 (2011).

\bibitem{KG} R.\ Kunjwal,  S.\ Ghosh, Phys. Rev. A 89, 042118 (2014).

\bibitem{abramskybranden} S. Abramsky and A. Brandenburger, New J. Phys. 13 113036 (2011).

\bibitem{cabelloncycle} M. Ara\'ujo, M.T. Quintino, C. Budroni, M.T. Cunha, and A. Cabello, Phys. Rev. A 88, 022118 (2013).

\bibitem{chavesfritz} R. Chaves and T. Fritz, Phys. Rev. A 85, 032113 (2012).

\bibitem{kunjspek}  R. Kunjwal and R.W. Spekkens, ``From Specker's scenario to noncontextuality inequalities," (in preparation), see also PIRSA:14010102,  http://pirsa.org/14010102,
for an earlier version.
\end{thebibliography}
\end{document}